\documentclass[twocolumn,showpacs,preprintnumbers,amsmath,amssymb]{revtex4}
\usepackage{epsfig}
\usepackage{graphicx}
\usepackage{textcomp}

\newcommand{\se}{s_{\rm ext}}
\newcommand{\0}{^{(0)}}
\newcommand{\1}{^{(1)}}

\newcommand{\Q}{{\cal Q}}

\newcommand{\pb}{{\bf p}}
\newcommand{\xb}{{\bf x}}
\newcommand{\rb}{{\bf r}}
\newcommand{\Xb}{{\bf X}}
\newcommand{\xbS}{{\bf x}_{\cal S}}

\newcommand{\pbS}{{\bf p}_{\cal S}}

\newcommand{\xbo}{{\bf x}_{\rm opt}}

\newcommand{\pbo}{{\bf p}_{\rm opt}}

\def\p12{p_{12}({\bf q},t)}

\begin{document}
\title{Speeding up  disease extinction with a limited amount of vaccine}

\author{M. Khasin$^{1}$, M.I. Dykman$^{1}$ and B. Meerson$^{2}$}

\affiliation{$^{1}$Department of Physics and Astronomy, Michigan State University, East Lansing, MI 48824 USA}

\affiliation{$^{2}$Racah Institute of Physics, Hebrew University
of Jerusalem, Jerusalem 91904, Israel}

\begin{abstract}
We consider optimal vaccination protocol where the vaccine is in short supply. In this case, disease extinction results from a large and rare fluctuation. We show that the probability of such fluctuation can be exponentially increased by vaccination. For periodic vaccination with fixed average rate, the optimal vaccination protocol is model independent and presents a sequence of short pulses. The effect of vaccination can be resonantly enhanced if the pulse period coincides with the characteristic period of the disease dynamics or its multiples. This resonant effect is illustrated using a simple epidemic model. If the system is periodically modulated, the pulses must be  synchronized with the modulation, whereas in the case of a wrong phase the vaccination can lead to a negative result. The analysis is based on the theory of fluctuation-induced population extinction in periodically modulated systems that we develop.

\end{abstract}


\maketitle
\section{Introduction}

Spreading of an infectious disease  is a random process. An important source of the randomness is demographic noise associated with the stochastic character of the events of infection, recovery, birth, death, etc. In a large population demographic noise is small on average, and the infection spread leads to an endemic state where a certain fraction of the population stays infected for a long time. The disease, however, can disappear as a result of a large rare fluctuation, an unlikely chain of events where, for example, susceptible individuals happen to avoid getting infected while infected ones recover \cite{Bartlett1960,Anderson,Andersson2000}. Then, if there is no influx of infected individuals from the outside, the population will be disease-free. Such spontaneous disappearance of a disease is an example of population extinction studied in stochastic population dynamics.

Spontaneous extinction is also important for various physical and chemical reaction systems. This is a consequence of the underlying similarity of the dynamics that result from short random events, like collisions between molecules that lead to chemical reactions and interactions between individuals that lead to the disease spread. Extinction can be understood for different types of systems within the same general formalism, which provides a broader scope for the present paper. Moreover, the method of optimal control of extinction that we propose can be applied to systems of various types.

A conventional way of fighting epidemics is via vaccination. If there is enough vaccine, the infection can be eradicated ``deterministically" by eliminating  the endemic state \cite{Scherer}.
The amount of available vaccine, however, is often insufficient. The vaccine may be expensive, or it  may be dangerous to store in large amounts, as in the case of anthrax, or it may be effectively short-lived due to mutations of the infection agent, as for HIV \cite{Robertson} and influenza \cite{hay}.

Even where the endemic state may not be eliminated deterministically, vaccination can dramatically affect the stochastic dynamics of the epidemics. The underlying mechanism is the change of the rate of large fluctuations leading to disease extinction. For a well mixed population, this rate $W_e$ is usually exponentially small for a large total population size $N\gg 1$, $W_e\propto \exp(-{\cal Q})$ with ${\cal Q}\propto N$,  \cite{Weiss1971,Leigh1981,Herwaarden1995,Andersson2000,Nasell2001,Elgart2004,Doering2005,Dykman2008,Kamenev2008}. We call ${\cal Q}$ the disease extinction barrier. Vaccination changes the value of ${\cal Q}/N$. In turn, this changes  the disease extinction rate exponentially strongly. This effect was previously discussed for vaccination applied at random \cite{Dykman2008}.

The goal of this paper is to find an optimal way of administering a limited amount of vaccine which would maximally increase the disease extinction rate. We find a vaccination protocol that applies for a broad class of epidemic models. Our approach is based on the observation that, in a large fluctuation that leads to disease extinction, the population is most likely to evolve in a well-defined way. It moves along the most probable path in the space of the dynamical variables which characterize different sub-populations \cite{Dykman2008,Kamenev2008,MS2009}. Vaccination perturbs the system as it moves along the optimal path. One can think of vaccination as ``force" and its effect as ``work" done on the system. This work reduces the barrier ${\cal Q}$. The problem then is to maximize the work for given constraints on the vaccine.

Optimization of the effect of vaccination resembles another problem of optimal control of random systems: controlling large fluctuations in noise-driven dynamical systems by applying an external field with a given average intensity \cite{Smelyanskiy1997,Vugmeister1997}.  There are, however, important differences, which come from the very nature of the control field. Indeed, vaccination only \textit{reduces} the number of susceptible individuals. In other words, as a control field, vaccination never changes sign. Then, as we find, if  the available amount of vaccine is constrained by a given mean vaccination rate, the optimal vaccination protocol turns out to be model-independent. This applies also to using a limited amount of medications and other situations where the control field drives the system only in one direction.

We assume that vaccination is applied periodically in time. In this case, the optimal protocol is to apply the vaccine as a sequence of $\delta$-like pulses. The disease extinction rate can strongly depend on the period of this sequence. Furthermore, the extinction rate can display exponentially sharp peaks when the pulse sequence period is close to the characteristic period of oscillations of the system in the absence of fluctuations, or to its multiples. We illustrate this resonant phenomenon for the Susceptible-Vaccinated-Infected-Recovered (SVIR) model.

Epidemics often display seasonal modulation \cite{Grassly}.  It is natural to apply a vaccine with period equal to the modulation period. As we show, there is a qualitative difference between the effect of a periodic vaccination in this case and in the case where seasonal modulation is absent. For a system with seasonal modulation, an improperly applied pulsed vaccination can actually \textit{reduce} the disease extinction rate and therefore prolong the duration of the epidemic. The overall effect of the pulsed vaccination critically depends here on the \textit{phase} at which the periodic pulses are applied.

The analysis of periodic vaccination, with and without seasonal variations, necessitates a general formulation of the extinction problem in periodically modulated stochastic populations. We extend the previous results for single-population systems \cite{Escudero2008,Assaf2008} to provide a complete extinction theory for multi-population systems in the eikonal approximation, and emphasize the distinction from the well-understood problem of switching between metastable states in periodically modulated systems with noise \cite{Ryvkine2006}.

Section II describes the class of epidemic models we consider in this work and develops an eikonal theory of disease extinction rate in periodically modulated systems.
Section III formulates the optimization problem for vaccination and presents its solution for a time-periodic vaccination in the limit of a small average vaccination rate. In Section IV we discuss the vaccination-induced reduction of the disease extinction barrier for two types of constraints on the vaccination period, a limited lifetime of the vaccine and a limited vaccine accumulation. Section V illustrates, on the example of the stochastic SVIR model, the phenomenon of resonant response to vaccination. Section V contains concluding remarks.

\section{The disease extinction rate}
\subsection{The model of population dynamics}
\label{sec:vaccine}
We consider stochastic disease dynamics in a well-mixed population which includes infected ($I$) and susceptible ($S$) individuals and possibly other population groups such as recovered or vaccinated.  The system state is described
by a vector $\Xb=(S,I,\ldots)$ with integer components equal to the sizes of different population groups. Along with $\Xb$ it is convenient to consider a quasi-continuous vector $\xb=\Xb/N$, where $N$ is the characteristic total population size, $N\gg 1$. We assume that the population dynamics is Markovian. It is quite generally described by the master equation for the probability distribution $P(\Xb,t)$,
\begin{eqnarray}
\label{eq:master_master}
\dot P({\bf X},t)&=&\sum_{\rb}\left[W({\bf X-r, r},t)P({\bf X-r},t)\right. \nonumber \\
&-&\left. W({\bf X, r},t)P({\bf X},t)\right].
\end{eqnarray}
Here $W(\Xb,\rb,t)$ is the rate of an elementary transition $\Xb \to \Xb + \rb$ in which the population size changes by $\rb=(r_1,r_2,\ldots)$. Examples of such transitions are infection of a susceptible individual as a result of contacting an infected individual, recovery of an infected individual or arrival of a susceptible individual.

We assume that there is no influx of infected individuals into the population. Therefore, there are no transitions from states where there are no infected to states where infected are present,
\begin{equation}
\label{eq:extinction_hyperplane}
W(\Xb,\rb,t)=0\qquad {\rm for}\quad X_E=0, \,r_E\neq 0,
\end{equation}
where  subscript $E$ is used for the component of $\Xb$ which enumerates infected, $X_E \equiv I$.

In the neglect of demographic noise the population dynamics can be described by the deterministic (mean-field)  equation for the  population size $\bar \xb$,
\begin{equation}
\label{eq:mean_field}
\dot{\bar\xb} = \sum_{\rb}\rb w(\bar\xb,\rb,t), \qquad  w(\xb,\rb,t)=W(\Xb,\rb,t)/N.
\end{equation}
It immediately follows from Eq.~(\ref{eq:master_master}) if the width of the probability distribution $P(\Xb,t)$ is set equal to zero.
\subsubsection{Stationary systems}
We start with the case where the transition rates $W(\Xb,\rb)$ are independent of time. An endemic state, where a finite fraction of population is infected for a long time, corresponds to an attracting fixed point $\xb_A$ of the dynamical system, Eq.~(\ref{eq:mean_field}). We will assume throughout this work that there is only one such  point. We will also assume that  Eq.~(\ref{eq:mean_field}) has one fixed point $\xbS$ in the hyperplane $x_E=0$. The state $\xbS$  is stable with respect to all variables except $x_E$. We call it the disease extinction state. If $x_E>0$ (there is a nonzero number of infected), the deterministic trajectory leaves the vicinity of $\xbS$ and approaches the endemic state $\xb_A$.

Due to demographic noise the endemic state is actually \textit{metastable}. A rare large fluctuation ultimately drives the population into a disease-free state. The most probable fluctuation brings the system to the fixed point  $\xbS$ \cite{Dykman2008,Kamenev2008}. The probability of such a fluctuation per unit time, \textit{i.e.}, the disease extinction rate $W_e$, is given by the probability current to $\xbS$, similarly to the problem of escape from a metastable state \cite{Kramers1940}. For time-independent $W(\Xb,\rb)$ this current is quasistationary for times $t_r\ll t\ll W_e^{-1}$, where $t_r$ is the characteristic relaxation time for the noise-free motion described by Eq.~(\ref{eq:mean_field}).

We note that, even though the state  $\xbS$ is a saddle point in the mean-field approximation, it differs from the  saddle-point states encountered in the problem of switching between metastable states of reaction systems. In the case of interstate switching, the rates of elementary transitions $W(\Xb,\rb)$ in the unstable direction are nonzero, and ultimately fluctuations drive the system away from the saddle point. In the case of extinction, fluctuations around $\xbS$ occur only in the extinction hyperplane, whereas the probability of exiting this hyperplane is zero.

If the system has, in the mean-field approximation, more than one steady state away from the extinction hyperplane, extinction can go in steps: from the endemic state to another steady state and, ultimately to the extinction hyperplane. In particular, if the only additional steady state is a saddle point at the boundary between the basins of attraction of $\xb_A$ and $\xbS$, the problem of extinction can be reduced to the problem of escape over this saddle point \cite{assaf09}.

\subsubsection{Periodically modulated systems}

The above picture can be extended to the case where  the transition rates are periodic functions of time, $W(\Xb,\rb,t+T)=W(\Xb,\rb,t)$. Periodicity of some of the rates in time is a natural way of modeling seasonal variations of epidemics \cite{Grassly}. The attracting solution of  Eqs.~(\ref{eq:mean_field}), which describes the endemic state, is no longer stationary. We will assume that this solution, $\xb_A(t)$, is periodic in time with the same period $T$, $\xb_A(t+T)=\xb_A(t)$. The asymptotic disease extinction state $\xbS(t)$ is also periodic in time; it lies in the hyperplane $x_E=0$.

The most probable fluctuation which causes extinction of the disease corresponds to a transition from $\xb_A(t)$ to $\xbS(t)$ \cite{Assaf2008}. An important characteristic of such a transition is the period-averaged disease extinction rate $W_e$. It can be introduced if the modulation period $T\ll W_e^{-1}$ and, in addition, $t_r\ll W_e^{-1}$. In this case, for time $t$ such that $t_r,T\ll t\ll W_e^{-1}$, a quasi-stationary time-periodic probability distribution is formed, centered at $\xb_A(t)$. The probability current from $\xb_A(t)$ to $\xbS(t)$ is also periodic in time, and the period-averaged value of this current gives $W_e$ \cite{Assaf2008}, in a direct analogy with the problem of switching between metastable states in noise-driven dynamical systems \cite{Smelyanskiy1999,Maier2001a,Lehmann2000c,Dykman2005a}.

\subsection{Eikonal approximation}
\subsubsection{Equations of motion}
We will be interested in evaluating the disease extinction barrier ${\cal Q}$ which gives the exponent in the disease extinction rate, $W_e\propto \exp(-{\cal Q})$, at $N \gg 1$. This barrier is entropic in nature, as it results from an unlikely sequence of elementary transitions. It can be found by either solving the mean first passage time problem for reaching $\xbS(t)$ \cite{Weiss1971,Leigh1981,Herwaarden1995} or by calculating the tail of the quasi-stationary probability distribution $P(\Xb,t)$ for $\xb$ close to $\xbS(t)$ \cite{Elgart2004,Dykman2008,Kamenev2008}. Here we choose the latter strategy and determine, to the leading order in $N$, the logarithm of the distribution tail. We seek the solution of Eq.~(\ref{eq:master_master}) in eikonal form, $P({\bf X},t)=\exp[-N s({\bf x},t)]$  \cite{Kubo73,Gang1987,Dykman1994d}. In the limit of large $N$, from Eq.~(\ref{eq:master_master}) we obtain the following equation for $s({\bf x},t)$:
\begin{eqnarray}
\label{eq:eikonal_approximation}
    &&\partial_t s=-H(\xb,\partial_{\xb}s,t),\\
    &&H(\xb,\pb,t)=\sum\nolimits_{\bf r}w(\xb,\rb,t)\left[\exp(\pb\rb)-1\right].\nonumber
\end{eqnarray}
Here, we have taken into account that, typically, $|\rb|\ll N$, and $W(\Xb,\rb,t)$ depends on $\Xb$ polynomially, whereas $P$ is exponential in $\Xb$. Therefore we expanded $P({\bf X+r},t)\approx P({\bf X},t)\exp(-\rb\partial_{\xb}S)$ and replaced, to the leading order in $1/N$, $w(\xb-N^{-1}\rb,\rb,t)$ by $w(\xb,\rb,t)$.

Equation (\ref{eq:eikonal_approximation}) has the form of the Hamilton-Jacobi equation for an auxiliary Hamiltonian system with Hamiltonian $H(\xb,\pb,t)$; $s(\xb,t)$ is the action of this system. The Hamilton equations of motion are
\begin{eqnarray}
\label{eq:eom_Hamiltonian}
&&\dot\xb=\sum\nolimits_{\rb}\rb w(\xb,\rb,t)e^{\pb\rb}, \nonumber\\
&&\dot \pb=-\sum\nolimits_{\rb}\partial_{\xb} w(\xb,\rb,t)\left(e^{\pb\rb}-1\right).
\end{eqnarray}
These trajectories determine, in turn, the most probable, or optimal, trajectories that the system follows in a fluctuation to a given state $\xb$ at time $t$. We will calculate action $s(\xb,t)$ using these trajectories and thus find the exponent in the distribution $P(\Xb,t)$.

\subsubsection{Boundary conditions for the optimal extinction trajectory}
To find the boundary conditions for Hamiltonian trajectories (\ref{eq:eom_Hamiltonian}), we note that the quasi-stationary distribution $P(\Xb,t)$ has a Gaussian maximum at $\Xb_A(t)$.
This means that, close to attractor  $\xb_A(t)$, action $s(\xb,t)$ is quadratic in $|\xb-\xb_A)|$ for stationary systems, whereas for  periodically modulated systems $s(\xb,t)= s(\xb,t+T)$ is quadratic in the distance from trajectory $\xb_A(t)$ \cite{Dykman1993}. On the Hamiltonian trajectories that give such action, the momentum $\pb \equiv \partial_{\xb}s \to 0$ for $\xb\to \xb_A(t)$, and since $\xb= \xb_A(t), \pb=0$ is a fixed point (a periodic trajectory) of the Hamiltonian dynamics, the trajectories of interest start at $t \to -\infty$,
\begin{eqnarray}
\label{eq:Lagrangian_defined}
&&s(\xb,t)=\int_{-\infty}^tdt\,L(\dot\xb,\xb,t),\\
&&L(\dot\xb,\xb,t)=\sum\nolimits_{\bf r}w(\xb,\rb,t)\left[(\pb\rb-1)e^{\pb\rb} + 1\right].\nonumber
\end{eqnarray}
In the Lagrangian $L$, Eq.~(\ref{eq:Lagrangian_defined}), $\pb$ should be expressed in terms of $\dot\xb,\xb$ using Eq.~(\ref{eq:eom_Hamiltonian}). Since $w\geq 0$, we have $L \geq 0$.

Extinction barrier $\cal Q$ is determined by $N s({\bf x},t)$ for $\xb$ in the extinction hyperplane, $x_E=0$. In the spirit of the eikonal approximation, we have to find such $(\xb,t)$ in this hyperplane that $s({\bf x},t)$ be minimal. The minimum determines the boundary conditions for the optimal Hamiltonian trajectory of extinction, $(\xbo(t), \pbo(t))$. The condition that $s({\bf x},t)$ is minimal with respect to ${x}_{i\neq E}$ on the extinction hyperplane means that ${p}_{i}=\partial_{x_i}s\to 0$ for $i\neq E$ as the trajectory $(\xbo(t), \pbo(t))$ approaches the hyperplane.
The minimum of $s({\bf x},t)$  with respect to $t$ within the period of modulation is reached if $H(\xb,\pb,t)\to 0$ as the trajectory  approaches the hyperplane.

A consequence of conditions $H(\xb,\pb,t)\to 0$ and $p_{i\neq E}\to 0$ is that the momentum component $p_E$ remains bounded on trajectory $(\xbo(t), \pbo(t))$. Indeed, near the extinction hyperplane, $x_E \ll 1$, we have
\begin{eqnarray}
\label{eq:approach}
\dot{x}_{E}&=&\sum\nolimits_{\rb}r_E  w(\xb,\rb,t) e^{\pb \rb}\nonumber\\
 &&\approx x_E \sum\nolimits_{\rb}r_E  [\partial w(\xb,\rb,t)/\partial x_E]_{x_E=0} e^{p_E r_E}.
\end{eqnarray}
Here, we assumed that $w(\xb,\rb,t)$ is nonsingular at $x_E\to 0$ and, since $w=0$ for $x_E=0$ and $r_E\neq 0$ [cf. Eq.~(\ref{eq:extinction_hyperplane})], we expanded $w$ in $x_E$ to the lowest order. Let us assume now that $|p_E| \to \infty$ for $x_E\to 0$. Then only the term with maximal $-r_E\equiv -r_{Em}$ should be kept in the sum over $r_E$ in Eq.~(\ref{eq:approach}); it is also clear that $p_E$ should be negative, otherwise the trajectory would not approach $x_E=0$. From the Hamilton equation for $p_E$ and Eq.~(\ref{eq:approach}) it follows that $dp_E/dx_E\approx -1/x_E r_{Em}$. This relation, along with the explicit form of the Hamiltonian $H$, show that, if $p_E$ were diverging for $x_E\to 0$, the Hamiltonian would not become equal to zero but would remain $\approx \partial w/\partial x_E$ with the derivative calculated for $x_E= 0$ and $r_E=r_{Em}$. This contradiction shows that the assumption $|p_E|\to \infty$ is wrong, $p_E$ remains limited for $x_E\to 0$.

Equation(\ref{eq:approach}) shows that $x_E\to 0$  exponentially as $t\to \infty$. As $x_E$ approaches zero, variables $x_{i\neq E}$ are approaching the equilibrium position in the hyperplane $x_E=0$. This happens because ${p}_{i\neq E}\to 0$ and the dynamics of $x_{i\neq E}$ in the hyperplane is described by the mean-field equations, Eq.~(\ref{eq:mean_field}). Therefore,
\begin{eqnarray}
\label{eq:barrier_in_terms_of_s}
&&{\cal Q}=N\se, \qquad \se=\int_{-\infty}^{\infty}dt L(\dot\xb,\xb,t),\\
&&\xb(t)\to \xbS(t), \;\pb(t) \to \pbS(t) \;{\rm for}\; t\to \infty.\nonumber
\end{eqnarray}
Function $\pbS(t)$ is periodic in time, with $p_{i\neq E}=0$ and with hitherto unknown component $p_E(t)$, which is discussed below.

The optimal trajectory  $(\xbo(t),\pbo(t))$ is a heteroclinic Hamiltonian trajectory that goes from periodic orbit
$(\xb_A(t),\pb={\bf 0})$ to periodic orbit $(\xbS(t),\pb_S)$, and the action for extinction $\se$ is calculated along this trajectory. The trajectory $\xbo(t)$ is the optimal path to disease extinction: it describes the most probable sequence of elementary transitions leading to extinction. We note that, in periodically modulated systems, there is one optimal path per period, whereas in stationary systems trajectories $(\xbo(t),\pbo(t))$ are time-translation invariant.

\subsubsection{The $t\to\infty$ value of the momentum on the Hamilton trajectory}

The momentum component $p_E$ is generically nonzero, as found for stationary systems \cite{Herwaarden1995, Dykman2008, Kamenev2008, Khasin2009}. For periodically modulated systems, one can show that $p_E\neq 0$ by extending  the arguments presented in Ref.~\cite{Dykman2008,Khasin2009}. This amounts to showing that the stable manifold of the periodic orbit $(\xbS(t),\pb={\bf 0})$ lies entirely in the invariant  hyperplane $x_E=0$, $p_{i\neq E}= 0$, and, as a consequence,  does not intersect  the unstable manifold of the periodic orbit $(\xb_A(t),\pb={\bf 0})$ . Such intersection is necessary in order to have a heteroclinic trajectory that would go from $(\xb_A(t),\pb={\bf 0})$ to  $(\xbS(t),\pb={\bf 0})$.

The  hyperplane $x_E=0$, $p_{i\neq E}= 0$  is formed by trajectories
\begin{eqnarray}
\label{eq:dyn}
\dot{x}_{i\neq E}&=&\sum\nolimits_{\rb} \left[w(\xb,\rb,t)\right]_{x_E=0} r_i, \\
\dot{p}_{E}&=&-\sum\nolimits_{\rb} \left[\partial_{x_E} w(\xb,\rb,t)\right]_{x_E=0} \left(e^{p_E r_E }-1\right). \nonumber
\end{eqnarray}
The invariance of this hyperplane is a consequence of Eq.~(\ref{eq:extinction_hyperplane}), which leads to $\dot{p}_{i\neq E}=0$ and $\dot{x}_E=0$ for $p_{i\neq E}=0$ and $x_E=0$.

To prove that the stable manifold of  $(\xbS(t),\pb={\bf 0})$ lies entirely in the invariant hyperplane $x_E=0$, $p_{i\neq E}= 0$, we first show that the trajectories, which are described by Eq.~(\ref{eq:dyn}) and which start close to the state $(\xbS(t),\pb={\bf 0})$, approach this state for $t \to \infty$. Then, since the dimension of the hyperplane $x_E=0$, $p_{i\neq E}= 0$ is equal to the dimension of the stable manifold of $(\xbS(t),\pb={\bf 0})$, we conclude that the stable manifold indeed lies in the hyperplane.

Equations (\ref{eq:dyn}) for ${x}_{i\neq E}$ are the  mean-field equations in the extinction hyperplane $x_E=0$, cf. Eqs.~(\ref{eq:mean_field}), and therefore
${x}_{i}\to (\xbS(t))_i$ for $t \to \infty$. Linearization of Eq.~(\ref{eq:dyn}) for $p_E$ about $(\xbS(t),\pb={\bf 0})$ gives
\begin{eqnarray}
\label{eq:momentum}
\dot{p}_{E}&=&-\sum\nolimits_{\rb} \left[\partial_{x_E} w(\xb,\rb,t)\right]_{\xbS(t)}p_E r_E.
\end{eqnarray}
We compare this equation with the mean-field equation for $x_E$ near $\xbS(t)$. The latter has the form $\dot{x}_{E}= x_E \sum\nolimits_{\rb}r_E  [\partial_{x_E}w(\xb,\rb,t)]_{\xbS(t)}$.
Since the state $\xbS(t)$ is  unstable in  $x_E$ direction in the mean-field approximation, from Eq.~(\ref{eq:momentum}) $\dot{p}_{E}/{p}_{E}<0$. Therefore, all trajectories on the hyperplane $x_E=0$, $p_{i\neq E}= 0$ close to the state $(\xbS(t),\pb={\bf 0})$ approach this state  asymptotically as $t \to \infty$, and thus the stable  manifold  of $(\xbS(t),\pb={\bf 0})$ lies in this hyperplane.

From the above analysis one concludes that there are no Hamiltonian trajectories that would go from $(\xb_A(t),\pb={\bf 0})$ to $\bigl(\xbS(t),\pb={\bf 0}\bigr)$. Therefore the optimal trajectory leading to extinction should go to a state $\bigl(\xbS(t),[\pbS(t)]_E\bigr)$ with $[\pbS(t)]_E \neq 0$.

\section{Optimal vaccination}

\subsection{Constraint on the vaccination protocol}
\label{sec:FormulationOfTheVariationalProblem}

Vaccination increases the number of individuals who are at least temporarily immune to the disease. It thus reduces the pool of susceptible individuals and ultimately leads to a reduction of the number of infected. When the available amount of vaccine is small, so that the disease extinction still requires a large fluctuation, the goal of vaccination is to reduce the disease extinction barrier $\Q$.

An outcome of vaccination is often modeled as the creation of a sub-population of  vaccinated individuals out of susceptibles. The corresponding elementary transition rate is $W(\Xb,\rb)=\xi_0(t)X_S$ for $r_S=-1$, $r_V=1$ and $r_{i\neq S,V}=0$, where subscripts  $V$ and $S$ refer to  vaccinated and susceptible individuals, respectively, and $\xi_0(t)$ is the control field that characterizes the vaccination (subscript  $S$ should not be confused with subscript $\cal{S}$ used to indicate the extinction state). Another model is vaccination of newly arrived susceptibles, which leads to an effective reduction of the arrival rate $\mu N$. In this model, the elementary transition rate for the arrival is  $W(\Xb,\rb)=N[\mu-\xi_0(t)]$ for $r_S=1$ and $r_{i\neq S}=0$, with $\xi_0(t)N$ being the change in the arrival rate due to vaccination.

We will consider a general model where vaccination modifies the rate of an elementary transition of a certain type; the change of the population in the corresponding transition is $\rb_{\xi}$. The field $\xi_0(t)$ characterizing the vaccination is assumed to be weak. The affected rate has the form $W(\Xb,\rb_{\xi},t)=W^{(0)}(\Xb,\rb_{\xi},t)+\xi_0(t)W^{(1)}(\Xb,\rb_{\xi},t)$, with $W^{(0)}$ being the rate without vaccination. The vaccination  either increases or decreases the rate, as for transitions from susceptibles to vaccinated or for vaccination of  newly arrived susceptibles, respectively. Therefore, we will assume without loss of generality that $\xi_0(t)\ge 0$ and that $W^{(1)}(\Xb,\rb_{\xi},t)$ is either  positive or  negative. We consider models in which the number of susceptibles changes by $1$ in an elementary transition associated with vaccination,  $(\rb_{\xi})_S=\pm 1$. We note that the analysis can be immediately extended to describe other processes, like modification of the infection rates \cite{Andersson2000} or recovery acceleration by administrating medicine.

It should be noted that the vaccination model adopted in this work is probabilistic by nature.  An alternative is where vaccination is done in a pre-determined fashion, when a certain number of individuals are vaccinated per unit time at a given time. The analysis of such deterministic  vaccination lies beyond the scope of this paper.

We will assume that vaccination is periodic, $\xi_0(t)=\xi_0(t+T)$, and that the amount of vaccine available per period $T$ is limited. We model this limitation as a constraint on the ensemble-averaged number of individuals vaccinated per period $T$. The constraint can be written as
\begin{eqnarray}
\label{eq:mean_vaccine_rate}
&&T^{-1} \int_0^Tdt\,  \xi_0(t) \sum_{\Xb} \left|W^{(1)}(\Xb,\rb_{\xi},t)\right|P(\Xb,\rb,t) \nonumber\\
&& =N\Xi.
\end{eqnarray}
Here, $\Xi$ is the average vaccination rate  rescaled by the characteristic population size $N$. The constraint  is  well-defined for $t_r \ll t \ll W_e^{-1}$, where $P(\Xb,\rb,t+T)\approx P(\Xb,\rb,t)$. Since for $N \gg 1$ the population distribution sharply peaks at the endemic state $\Xb_A(t)$, the sum over $\mathbf{X}$ in Eq.~(\ref{eq:mean_vaccine_rate}) can be replaced by  $\left|W^{(1)}(\Xb_A(t),\rb_{\xi},t)\right|$, in the leading order in $1/N$.

In the presence of vaccination, one can still seek a solution of the master equation in the eikonal form. The exponent $\Q$ in the extinction rate is again given by the action of an auxiliary Hamiltonian system, Eq.~(\ref{eq:barrier_in_terms_of_s}). The Hamiltonian now has the form
\begin{eqnarray}
\label{eq:H_0_and_H_1}
&&H(\xb,\pb)=H^{(0)}(\xb,\pb) + \xi_0(t)H^{(1)}(\xb,\pb), \nonumber \\
&&H^{(0)}(\xb,\pb)= \sum\nolimits_{\rb}w^{(0)}(\xb,\rb,t)(e^{\pb\rb}-1), \\
&&H^{(1)}(\xb,\pb)=w^{(1)}(\xb,\rb_{\xi},t)(e^{\pb\rb_{\xi}}-1). \nonumber
\end{eqnarray}
Our goal is to find the optimal form of  $\xi_0(t)$ which would minimize the disease extinction barrier $\Q$ subject to constraint (\ref{eq:mean_vaccine_rate}). Since $w^{(1)}(\xb_A(t),\rb_{\xi},t)$ is a known periodic function of time, we can equivalently search for the optimal \textit{vaccination rate} $\xi(t)\equiv \xi_0(t)\left|w^{(1)}(\xb_A(t),\rb_{\xi},t)\right|$. It minimizes the functional
\begin{eqnarray}
\label{eq:var_func}
&&\tilde s_{\rm ext}\left[\xi(t)\right]=\se\left[\xi(t)\right]+\lambda T^{-1}\int\nolimits_0^T \left[\xi(t)-\Xi\right]dt, \\
&&\xi(t)=\xi(t+T)=\xi_0(t)\left|w^{(1)}(\xb_A(t),\rb_{\xi},t)\right| \geq 0, \nonumber
\end{eqnarray}
where $\lambda$ is the Lagrange multiplier.  The functional $\se[\xi]$ is given by Eq.~(\ref{eq:barrier_in_terms_of_s}), where the Lagrangian corresponds to the Hamiltonian (\ref{eq:H_0_and_H_1}) and depends on the vaccination rate $\xi(t)$.

\subsection{Vaccination protocol for stationary systems}
\subsubsection{Double optimization problem}

For a low average vaccination rate $\Xi$ it suffices to keep in the action $\se[\xi(t)]$ only the leading-order term in $\xi(t)$. Since Hamiltonian (\ref{eq:H_0_and_H_1}) is linear in $\xi$, this term is linear in $\xi$, too. In the spirit of the standard perturbation theory for Hamiltonian systems \cite{LL_Mechanics2004}, the change in the action, caused by the small perturbation, can be calculated along the unperturbed trajectory $(\xbo\0(t),\pbo\0(t))$ of the Hamiltonian $H\0$, which describes the optimal path of disease extinction in the absence of vaccination.

We start with the case of systems, which are stationary in the absence of vaccination. For such systems
\begin{eqnarray}
\label{eq:linear_approx}
&&\se[\xi(t)]=\se\0 + \se\1[\xi(t)], \\
&&\se\1[\xi(t)]=\min_{t_0}\int\nolimits_{-\infty}^{\infty}dt \chi(t-t_0)\xi(t),\nonumber\\
&&\chi(t) = -H\1\bigl(\xbo\0(t),\pbo\0(t)\bigr)\left|w^{(1)}(\xb_A,\rb_{\xi})\right|^{-1}. \nonumber
\end{eqnarray}
The quantity $\chi(t)$ is called logarithmic susceptibility \cite{Smelyanskiy1997c,Smelyanskiy1999a,Dykman2008,Assaf2008}; it gives the change of the logarithm of the extinction rate, which is linear in the vaccination rate for moderately low vaccination rate.

The minimization over $t_0$ in Eq.~(\ref{eq:linear_approx}) accounts for lifting  the time-translational invariance of the optimal extinction paths mentioned earlier.
For $\xi(t)\equiv 0$,  extinction can occur at any time ($t_r\ll t\ll W^{-1}$) with rate $W_e$. Periodic vaccination synchronizes extinction events; it periodically modulates the extinction rate, and the modulation is exponentially strong for $N|s^{(1)}_{\rm ext}|\gg 1$ (see below). Formally, in a modulated system there is only one optimal extinction path per period, as explained in Sec.~II, which is here reflected in the minimization over $t_0$. This optimal path minimizes  the disease extinction barrier $\Q=N\se$ \cite{Smelyanskiy1997c,Smelyanskiy1999a,Escudero2008,Assaf2008}.  Equation ~(\ref{eq:linear_approx}) is closely related to the Mel'nikov theorem for dynamical systems \cite{Guckenheimer1987, Assaf2008}.

The constraint for minimizing the action over $\xi(t)$ in Eq.~(\ref{eq:var_func}) has a form of an integral over the vaccination period $T$. It is therefore convenient to write action $\se\1$ also in the form of such an integral,
\begin{eqnarray}
\label{eq:s1_periodic}
&&\se\1[\xi(t)]= \min_{t_0}\int_0^Tdt\xi(t)\chi_T(t-t_0),\nonumber\\
&&\chi_T(t)=\sum_{n=-\infty}^{\infty}\chi(t+ n T).
\end{eqnarray}
The function $\chi_T(t)$ is obtained by superimposing the parts of $\chi(t)$ which differ by $T$. By construction, $\chi_T(t)$  is periodic in time $t$.

\subsubsection{Temporal shape of optimal vaccination}

To find the optimal shape of vaccination rate $\xi(t)$ we  first minimize the time integral in the variational problem Eqs.~(\ref{eq:var_func}) -- (\ref{eq:s1_periodic}) with respect to $\xi(t)$ for a given $t_0$. Since $\xi(t)\geq 0$, it is convenient to perform the minimization with respect to $\xi^{1/2}(t)$ rather than $\xi(t)$. The minimization shows that $\xi^{1/2}(t)\neq 0$ only for $t=t_{\lambda}$, where $t_{\lambda}$ is given by equation $\chi_T(t_{\lambda}-t_0)=-\lambda/T$. From the constraint on the period-averaged $\xi(t)$ we then have
\begin{equation}
\label{eq:ci_explicit}
\xi(t)=\Xi \,T\sum_n \delta (t-t_{\lambda}+nT).
\end{equation}
Substituting this expression into the functional $\tilde s_{\rm ext}$ and minimizing with respect to $t_0$, we obtain the action in a simple explicit form
\begin{eqnarray}
\label{eq:action_minimized}
&&\se=\min\tilde s_{\rm ext}=\se\0+\se\1,\nonumber\\
&&\se\1= \Xi \,T\min_{0\leq t < T}\chi_T(t).
\end{eqnarray}
Alternatively, this expression can be rewritten in terms of the Fourier transform of the logarithmic susceptibility:
\begin{eqnarray}
\label{eq:chi_Fourier}
&&\se\1 = \Xi \,\min_t\sum\nolimits_n\tilde\chi(n\Omega)\exp[in\Omega t],\\
&&\tilde\chi(\omega)=\int\nolimits_{-\infty}^{\infty}dt \chi(t)\exp(i\omega t),\nonumber
\end{eqnarray}
where $\Omega=2\pi/T$ is the cyclic  frequency of vaccination.

We are interested in the solution for which $\se\1$ is negative, which requires $\min\chi_T(t) < 0$. Only in this case will vaccination reduce the barrier for disease extinction and increase the disease extinction rate. The barrier reduction due to vaccination, $\Q\1=N\se\1\propto N\Xi$, becomes large for $N\gg 1$ even if the average vaccination rate $\Xi$ is small. Thus, for not too small vaccination rates, where the eikonal approximation is valid \cite{Smelyanskiy1997c, Assaf2008}, the effect of vaccination on the disease extinction rate is exponentially strong.

The expression for the  action change $\se\1$, Eq.~(\ref{eq:action_minimized}), can be also obtained in a more intuitive way. Indeed, since  $\xi(t)$  is non-negative, it follows from Eq.~(\ref{eq:s1_periodic}) that
\begin{eqnarray}
\label{eq:qualitative_s1_minimization}
\se\1[\xi(t)] \geq \min_{t}\chi_T(t)\int\nolimits_0^T dt\xi(t)  = \Xi T \min_{t}\chi_T(t).
\end{eqnarray}
The minimum is provided by $\xi(t)=\Xi \,T\sum_n \delta (t-t_{min}+nT)$. Formally, $t_{min}$ is the instance of time where $\chi_T(t)$ is minimal. In fact, it is the optimal path that adjusts to the vaccination pulses so as to increase the probability of disease extinction. This provides the mechanism of synchronization by vaccination. Equation~(\ref{eq:qualitative_s1_minimization}) immediately leads to Eqs.~(\ref{eq:ci_explicit}) and (\ref{eq:action_minimized}) with $t_{\lambda}$ replaced by $t_{min}$.

In addition to the constraint on the average vaccination rate, there may be an upper limit on the instantaneous vaccination rate, which is imposed by condition $w(\xb,\rb)=w^{(0)}(\xb,\rb_{\xi})+\xi(t)w^{(1)}(\xb,\rb_{\xi})\ge 0$. In the case where $w^{(1)}(\xb,\rb_{\xi},t)<0$, as for vaccination of  newly arrived susceptibles, this condition is met provided $\xi_0(t) \le \xi_{0m}\equiv \min \{w^{(0)}(\xb,\rb_{\xi})/\left|w^{(1)}(\xb,\rb_{\xi})\right|\}$. In this case the optimal vaccination protocol changes.

The new protocol can be found from the variational problem (\ref{eq:var_func}) by changing from $\xi(t)\equiv \xi_0(t)\left|w^{(1)}(\xb_A,\rb_{\xi})\right|$ to an auxiliary function $\eta(t)$ such that
$\xi_0(t)=\xi_{0m}[1+\eta^2(t)]^{-1}$, and then finding the minimum of $\tilde s_{\rm ext}$ with respect to $\eta(t)$. This choice satisfies the constraints $0\leq \xi_0(t)\leq \xi_{0m}$. Variation with respect to $\eta(t)$ shows that $\tilde s_{\rm ext}$ has an extremum for $\eta(t)=0$ or $\eta(t)=\infty$ for $t \neq t_{\lambda}$, where $t_{\lambda}$ is given by equation $\chi_T(t_{\lambda}-t_0)=-\lambda/T$. The value of $\eta(t)$ at the isolated instances $t = t_{\lambda}$ is arbitrary. Only regions where $\eta(t)=0$, so that $\xi_0(t) = \xi_{0m}$, contribute to $\tilde s_{\rm ext}$. Obviously, $\tilde s_{\rm ext}$ is minimal when $\xi_0(t) = \xi_{0m}$ for $|t-t_{min}|\le \Delta t/2$, where  $t_{min}$ is the time when $\chi_T(t)$ is minimal and $\Delta t$ is determined by the average vaccination rate $\Xi$. In other words, the vaccination rate $\xi(t)$ has the form of periodic rectangular pulses of width $\Delta t$, centered at $t_{min}+nT$, $n=0,\pm 1, \pm 2,...$. The pulse width is
\begin{eqnarray}
\label{eq:deltat}
\Delta t = \frac{\Xi T }{\xi_{0m}\left|w^{(1)}(\xb_A,\rb_{\xi})\right|}.
\end{eqnarray}
Since the vaccination rate is limited by the rate of elementary transitions without vaccine, we have $\xi_{0m}\left|w^{(1)}(\xb_A,\rb)\right| \lesssim t^{-1}_r$. Then for weak vaccination, $\Xi T \ll 1$, from Eq.~(\ref{eq:deltat}) $\Delta t \ll t_r$.  Therefore, $\chi_T(t)=\chi_T(t_{min})$  during the pulse of $\xi(t)$, to the leading order in $\Xi T$ [we note that $\chi_T(t)$ may vary on a time scale shorter than $t_r$, see below; however, this time scale is always long compared to $\Delta t$ for sufficiently weak modulation]. The resulting change of the action is again given by Eq.(\ref{eq:action_minimized}).

\subsection{Vaccination protocol for periodically modulated systems}

Optimal vaccination in periodically modulated systems requires a separate consideration.  Here, there is only one optimal extinction path per period $T$  in the absence of vaccination. When the average vaccination rate $\Xi$ is low, vaccination with the same period $T$ will only weakly perturb this path.  To first order in $\Xi$, the linear in $\xi$ term in the action still has the form of Eq.~(\ref{eq:s1_periodic}), but without minimization over $t_0$. Since $\xi(t)\geq 0$, the minimum of action is still achieved for $\xi(t)=\Xi \,T\sum_n \delta (t-t_{min}+nT)$, but now $t_{\min}$ is uniquely determined by the strong modulation. In other words, the modulation uniquely determines the phase of the optimal vaccination pulses. The resulting expression for $\se\1$ for the optimal vaccination protocol has the form of Eq.~(\ref{eq:action_minimized}). If the vaccination pulses are applied at a wrong time, \textit{i.e.} if the  phase difference between the vaccination and the modulation differs from the optimal one, the vaccination will be not as efficient and may even be harmful: it may prolong the lifetime of the endemic state by \textit{increasing} the disease extinction barrier $\Q$.

\section{Disease extinction barrier as a function of vaccination period}

The vaccination-induced reduction of the disease extinction barrier $\Q\1=N\se\1$, as given by Eqs.~(\ref{eq:action_minimized}), depends on the interrelation between the vaccination period $T$ and the characteristic time scales of the logarithmic susceptibility $\chi(t)$. Function $\chi(t)$ may or may not  oscillate in time, but generally $\chi(t)$ is relatively large within a time interval of the order of the relaxation time of the system $t_r$ \cite{Smelyanskiy1997c,Assaf2008}. To reveal some qualitative features of the effect of vaccination and in particular, its dependence on the vaccination period, we  will consider $\se\1$  for two types of constraint on this period.

\subsubsection{Limited lifetime of the vaccine}

The vaccination period $T$ is  naturally limited by the effective lifetime $\tau_{\texttt{v}}$ of the vaccine. This lifetime is usually determined by the maximum storage time of the vaccine and/or by the mutation rates of the infectious agent. If $\tau_{\texttt{v}}$ is long, $\tau_{\texttt{v}}\gg t_r$,  vaccination can be made most efficient by increasing the vaccination period up to $\sim \tau_{\texttt{v}}$. Indeed, as it follows from Eq.~(\ref{eq:action_minimized}), $\se\1\propto T$ in this case. This result is easy to understand. Even though a decrease of the vaccine pulse frequency $\Omega=2\pi/T$ causes a decrease of the prefactor in the disease extinction rate $W_e$, the exponential factor $\exp(-N\se\1)$ in $W_e$ increases sharply. Indeed, it can be seen from Eqs.~(\ref{eq:eom_Hamiltonian}), (\ref{eq:H_0_and_H_1}) and (\ref{eq:linear_approx}) that, as the system moves along the optimal path to extinction, $\chi(t)$  is significant when the system is far from the stationary states $\xb_A$ and $\xbS$. The characteristic time scale of this motion is $\sim t_r$. For $T \gg t_r$ we have $\min\nolimits_{0\leq t\leq T} \chi_T(t)\approx \min\nolimits_t \chi(t)$ and
\begin{equation}
\label{eq:large_T}
\se\1=\Xi T\min_t\chi(t),\qquad \tau_{\texttt{v}}\gtrsim T\gg t_r.
\end{equation}

In the opposite limit of $\tau_{\texttt{v}}\ll t_r$, and thus $T\ll  t_r$, we have from Eq.~(\ref{eq:chi_Fourier})
\begin{equation}
\label{eq:small_T}
\se\1=\Xi \tilde\chi(0),\qquad t_r\gg\tau_{\texttt{v}}\gtrsim T,
\end{equation}
In this case the vaccination-induced reduction of the extinction barrier is independent of the vaccination period and is determined by the zero-frequency component of the logarithmic susceptibility.

An interesting situation may occur in the intermediate range $\tau_{\texttt{v}}\sim t_r$ if, in the mean-field description, the system approaches the endemic state in an oscillatory manner. In this case function $\chi(t)$ is also expected to oscillate. The oscillations are well-pronounced if their typical frequency is $\omega_0 \gg t_r^{-1}$.  It is clear from Eq.~(\ref{eq:chi_Fourier}) that a strong effect on disease extinction can be achieved by tuning the vaccination frequency $\Omega=2\pi/T$ or its overtones in resonance with $\omega_0$. An example of such a resonance will be discussed in Sec.~V.

\subsubsection{Limited vaccine accumulation}

A different situation occurs if the total amount  of vaccine that can be accumulated  is limited. This limitation implies that $\Xi T \leq M$ (note that  $M$ is the limit on the \textit{ensemble-averaged} amount of the accumulated vaccine). Such a constraint is typical for live vaccines, as it may be dangerous to store too much vaccine in this case. The actual average vaccination rate in this case is now $T$-dependent. We use the notation $\Xi_a$ for this rate, with $\Xi_a=\min(\Xi,M/T)$. This is $\Xi_a$ that should be used now in Eqs.~(\ref{eq:large_T}) and (\ref{eq:small_T}) for $\se\1$ in the limits $T\gg t_r$ and $T\ll t_r$, respectively.
 
The behavior of $\se\1$ with varying vaccination period $T$ depends on the form of the logarithmic susceptibility $\chi(t)$.  Let us first consider the case where $\chi(t)$ has a single local minimum (at $t=t_*$), and $|\chi(t)|$ monotonically decays to zero with increasing $|t-t_*|$. Here, once the vaccine accumulation has reached saturation with increasing $T$ (which happens for $\Xi T=M$), function $|\se\1|=M |\min\chi_T(t)|$ monotonically decreases with further increase in $T$. Indeed,
\begin{eqnarray}
\label{eq:single_mimimum_chi}
&&\frac{d}{dT}\min_{0<t<T}\chi_T(t) = \frac{\partial}{\partial T}\sum_n \chi(t_*-a_*T+nT) \nonumber\\
&&= \sum_n \left(\frac{d\chi(t-a_*T+nT)}{dt}\right)_{t=t_*}(n-a_*)> 0,
\end{eqnarray}
where $a_*$ gives the position of the minimum of $\chi_T(t)$ over $t$ and is given by equation $\partial\chi_T(t_*-a_*T)/\partial a_*=0$; we choose $0<a_*<1$ and take into account that $\min_{0<t<T}\chi_T(t)<0$. In Eq.~(\ref{eq:single_mimimum_chi}) we have used that, if $\chi(t)$ is minimal for $t=t_*$, then $d\chi/dt>0$ for $t>t_*$ and $d\chi/dt <0$ for $t<t_*$. It follows from Eq.~(\ref{eq:single_mimimum_chi}) that, once the vaccine accumulation has reached saturation, further increase in $T$ will only reduce the effect of the vaccine. This  result is understandable because, if $T$ increases beyond $M/\Xi$, the actual average vaccination rate $\Xi_a$ decreases.

A counterintuitive situation may occur if $\chi(t)$ is oscillating. Here the inequality (\ref{eq:single_mimimum_chi}) may be violated. As a result, the dependence of the effect of the vaccine on $T$ and, consequently, on the actual vaccination rate $\Xi_a$, may be nonmonotonic. An example of this behavior is discussed in the next section.

\section{Resonances in the stochastic SVIR model}
\label{sec:SVIR}
\begin{figure}[ht]
\includegraphics[width=3.4in]{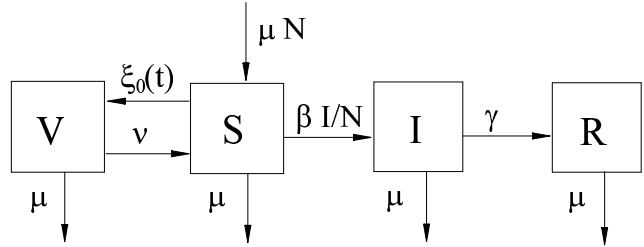}
\caption{The SVIR epidemic model with susceptible, vaccinated, infected and recovered sub-populations. The arrows indicate processes leading to changes of sub-population sizes; the corresponding rates are scaled per individual.}
\label{fig:SVIR_chart}
\end{figure}

We now apply some of our results to an important and widely used stochastic epidemic model, the Susceptibles-Vaccinated-Infected-Recovered (SVIR) model. The model is sketched in Fig.~\ref{fig:SVIR_chart}. In the absence of vaccination, $\xi_0(t)=0$, the SVIR model reduces to the stochastic SIR model with population turnover, which was originally introduced to describe the spread of measles, mumps, and
rubella, see \cite{Bartlett1960,Andersson2000}.  In the SIR model, susceptible individuals are brought in, individuals in all population groups leave (for example, die), a susceptible individual can become infected upon  contacting an infected individual, and an infected individual can recover. If we set $X_1=S, X_2= I$, and $X_3=R$, the rates of the corresponding processes are: (i) influx of the susceptibles, $W(\Xb,\rb)=\mu N$ for $r_1=1,r_{i\neq 1}=0$, (ii) leaving, with the same rate for all populations, $W(\Xb,\rb)=\mu X_i$ for $r_i=-1, r_{j\neq i}=0$, (iii) infection, $W(\Xb,\rb)=\beta X_1 X_2/N$ for $r_1=-1,r_2=1,r_{i\neq 1,2}=0$, and (iv) recovery of the infected, $W(\Xb,\rb)=\gamma X_2$ for $r_2=-1,r_3=1, r_{i\neq 2,3}=0$, Fig.~(\ref{fig:SVIR_chart}).

\begin{figure}[ht]
\includegraphics[width=3.4in]{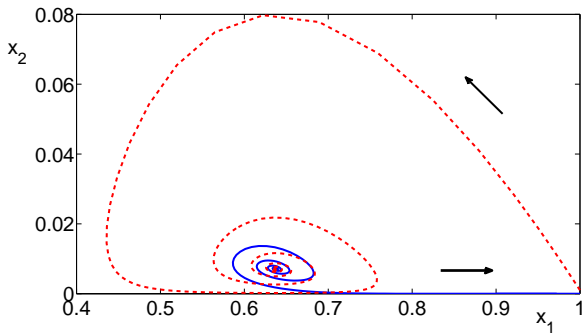}
\caption{The most probable trajectories in the stochastic SIR model on the plane of the scaled numbers of susceptibles and infected, $x_1=X_1/N$ and $x_2=X_2/N$, respectively. The dashed line shows a mean-field trajectory toward the endemic state, and the solid line shows the most probable trajectory followed during the fluctuation-induced disease extinction \cite{Kamenev2008}.  The plot refers to $\beta/\mu=80$,  and $\gamma/\mu=50$.}
\label{fig:SIR_oscillations}
\end{figure}

For $\beta>\Gamma\equiv\gamma+\mu$ the SIR model possesses a single endemic state. This state corresponds, in the mean-field theory, to an attracting fixed point on the two-dimensional phase plane of susceptibles and infected. At $\mu < 4 \,(\beta-\Gamma) (\Gamma/\beta)^2$ this attracting point is a focus. The populations of susceptibles, infected and recovered exhibit decaying oscillations in time  as the system approaches the endemic state. It was found in Ref.~\cite{Kamenev2008} that, in this parameter range, the populations oscillate also on the optimal disease extinction path. These oscillations are illustrated in Fig.~\ref{fig:SIR_oscillations}.

We will now incorporate vaccination and introduce a sub-population of vaccinated $X_4=V$. The vaccination is described by the transition rate  $W(\Xb,\rb)=\xi_0(t)x_1$ for $r_1=-1,r_4=1,r_{i\neq 1,4}=0$. The corresponding term in the Hamiltonian Eq.~(\ref{eq:H_0_and_H_1})  has the form $\xi_0(t)H\1$ with
\begin{equation}
\label{eq:H_1_vaccination_SVIR}
H\1(\xb,\pb)=x_1 \left(e^{p_4-p_1}-1\right).
\end{equation}
Vaccinated individuals leave at the same rate $\mu$ as individuals in other populations. For simplicity, we assume that the immunity from the vaccination is never lost. In this case fluctuations of the vaccinated population do not affect  fluctuations of other populations, and $p_4\equiv 0$ along the  optimal extinction path. Then from Eq.~(\ref{eq:linear_approx}), the logarithmic susceptibility is $\chi(t)=x_{1 \rm opt}^{(0)}(t)x_{1A}^{-1}\left(1-\exp[-p_{1 \rm opt}^{(0)}(t)]\right)$, where $x_{1 \rm opt}^{(0)}(t)$, $p_{1 \rm opt}^{(0)}(t)$ and $x_{1 A}$ are calculated for the SIR model.

\begin{figure}[ht]
\includegraphics[width=3.4in]{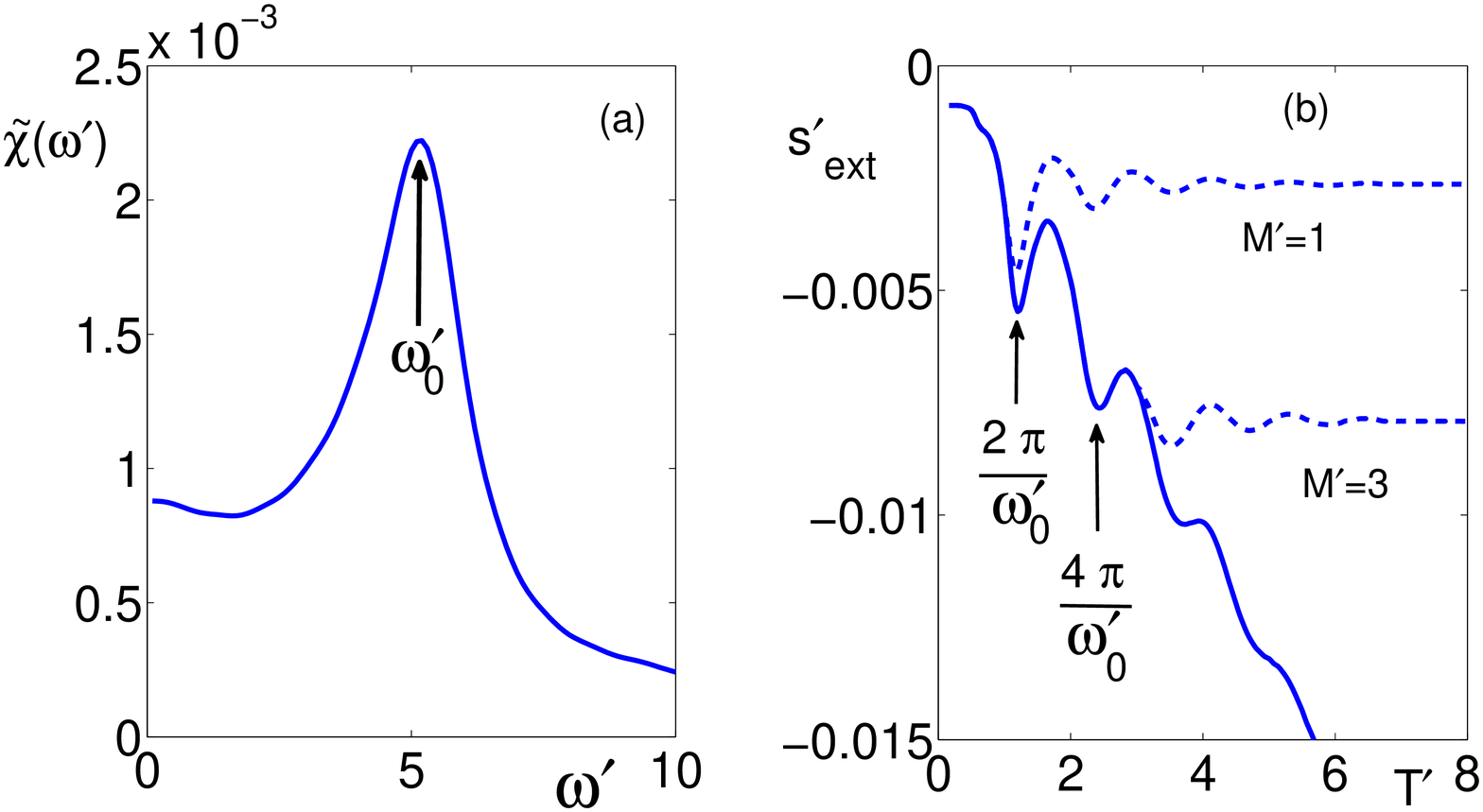}
\caption{(a) The Fourier transform of the logarithmic susceptibility  in the SIR model. The parameters are the same as in Fig.~\ref{fig:SIR_oscillations}; the rescaled frequency is $\omega^{\prime}=\omega/\mu$. The susceptibility spectrum displays a sharp peak at the characteristic vibration frequency $\omega_0$. (b) The change of the scaled extinction barrier $s_{\rm ext}^{\prime}=\mu \se\1/\Xi$ with vaccination period $T$. The solid line shows $s_{\rm ext}^{\prime}$ where there is no limit on vaccine accumulation, whereas the dashed lines refer to the case of limited accumulated vaccine amount. The accumulation limit $M$ is scaled by the small-$T$ average vaccination rate $\Xi$, $M'=\mu M/\Xi$, and $T^{\prime}=T\mu$. The locations of the resonances of $s_{\rm ext}^{\prime}$ are independent of $M$.}
\label{fig:resonances1}
\end{figure}

The Fourier spectrum of the logarithmic susceptibility $\tilde{\chi}(\omega)$ is plotted in Fig.~\ref{fig:resonances1}~(a). It corresponds to the optimal extinction path shown in Fig.~\ref{fig:SIR_oscillations}.  As one can see, the spectrum has a peak at the characteristic frequency of oscillations of the system in the absence
of vaccination $\omega_0$ (for the chosen parameter values $\omega_0 \approx 5.2\mu $).

We now consider the effect of the resonant peak in $\tilde{\chi}(\omega)$ on vaccination. The dependence of the scaled change of the disease extinction barrier $\se\1=\Q\1/N$ on vaccination period $T$ is shown in Fig.~\ref{fig:resonances1}~(b). The solid line in Fig.~\ref{fig:resonances1}~(b) shows the behavior of $\se\1$ where there is no limit on vaccine accumulation or, equivalently, for such periods where the limitation does not come into play and the actual vaccination rate $\Xi_a$ is independent of $T$. Function $|\se\1|\equiv -\se\1$ is seen to be strongly nonmonotonic, it displays pronounced maxima (which correspond to the minima of $\se\1$). They occur where the vaccination period $T$ coincides with the multiples of the characteristic period of the system motion without vaccination $2\pi/\omega_0$.

For limited vaccine accumulation $M$, the actual average vaccination rate depends on the vaccination period, $\Xi_a=\min(\Xi,M/T)$. Beyond a certain value of $T$, the increase of $T$ is accompanied by the decrease $\Xi_a$. This leads to a change of the dependence of $\se\1$ on $T$. Remarkably, $|\se\1|$ still displays resonant peaks at $2\pi n/\omega_0$ with integer $n$. Their amplitude decreases with increasing $n$. The occurrence of the peaks shows that, by tuning the vaccination period, the effect of the vaccination can be resonantly enhanced; the resonance in this case is in the exponent of the disease extinction rate, and therefore it is extremely strong. Counter-intuitively, since the actual average vaccination rate decreases with increasing $T$, a strong enhancement of the vaccine can be achieved where this rate is decreased. For example, in Fig.~\ref{fig:resonances1}~(b) the maxima of $|\se\1|$ for $\mu M/\Xi=1$ and $\mu M/\Xi=3$ are  achieved for $T$ in the range where $\Xi_a < \Xi$.

\section{Conclusions}
We have developed a theory of optimal periodic vaccination against an endemic disease for low average vaccination rate, as in the case where the vaccine is in short supply, or short lived, or cannot be stored in the sufficient amount. We assume that the vaccination rate is insufficient for eliminating the endemic state and thus exterminating the disease by ``brute force". However, vaccination can change the rate of disease extinction, which occurs spontaneously as a result of a comparatively rare fluctuation.  We show that the optimal vaccination leads to an exponentially strong increase of the disease extinction rate. This happens because the vaccine changes the effective entropic barrier that needs to be overcome for spontaneous extinction.

We find that the optimal vaccination protocol  is a periodic sequence of $\delta$-like pulses. This protocol is essentially model-independent, it only requires that the population be spatially uniform. In stationary systems, the phase of the pulses is irrelevant. In contrast, in periodically modulated systems, like in the case of seasonally varying infection, it is necessary to appropriately synchronize vaccination pulses with the modulation. Moreover, if the pulse phase is wrong, vaccination may hamper disease extinction.

For fixed average vaccination rate, the effect of vaccination in stationary systems increases with the increasing vaccination period. However, this increase is generally nonmonotonic and the disease extinction rate can display exponentially strong resonances. They occur if the vaccination period coincides with the period of decaying oscillations of the population, which characterize the approach to the endemic state in the mean-field (fluctuation-free) approximation. The resonances occur also where the vaccination period coincides with a multiple of the dynamical period. They are illustrated using the well-known SVIR model of population dynamics.

It turns out that, counterintuitively, the effect of vaccination can be sometimes enhanced by reducing the average vaccination rate. This happens where the mean-field dynamics is characterized by decaying oscillations and there is a constraint on the amount of vaccine that can be stored. In this case lowering the average vaccination rate can allow one to tune the vaccination period in resonance with the system dynamics. 

The analysis is based on the master equation for the population dynamics. We solve it in the eikonal approximation and reduce the problems of the tail of the distribution and of the extinction to Hamiltonian dynamics of an auxiliary system. A general formulation of the corresponding Hamiltonian problem is obtained for periodically modulated systems. The optimal vaccination protocol is found using this formulation with account taken of the constraint on the average vaccination rate. The feature of the problem that makes it different from other problems of optimal control of rare events is that the vaccination rate cannot be negative and it is the average vaccination rate that is given. The analysis can be extended to other problems of optimal control of fluctuation-driven extinction with similar constraints.

\subsection*{Acknowledgments}
The work at MSU was supported in part by the Army
Research Office and by NSF Grant PHY-0555346. B.~M. was supported by the US-Israel Binational Science Foundation Grant 2008075.


\end{document}